\documentclass[sigconf,nonacm]{acmart}
\usepackage{graphicx}
\usepackage{subcaption}
\usepackage{hyperref} 

\begin{document}


\title[ARENA: Energy Measurement Tool for Android Apps]{ARENA: A tool for measuring and analysing the energy efficiency of Android apps}

\author{Hina Anwar}
\orcid{0000-0002-4725-4636}
\affiliation{%
  \institution{University of Tartu}
  \city{Tartu}
  \country{Estonia}
}
\email{hina.anwar@ut.ee}

\renewcommand{\shortauthors}{Hina Anwar}

\begin{abstract} \label{section-ABSTRACT}
To build energy-efficient apps, there is a need to estimate and analyze their energy consumption in typical usage scenarios. The energy consumption of Android apps could be estimated via software-based and hardware-based approaches. Software-based approaches, while easier to implement, are not as accurate as hardware-based approaches. The process of measuring the energy consumption of an Android app via a hardware-based approach typically involves 1) setting up a measurement environment, 2) executing the app under test on a mobile device, 3) recording current/voltage data via a hardware device to measure energy consumption, and 4) cleaning and aggregating data for analyses, reports, and visualizations. Specialized scripts are written for selected hardware and software components to ensure reliable energy measurements. The energy measurement process is repeated many times and aggregated to remove noise. These steps make the hardware-based energy measurement process time-consuming and not easy to adapt or reproduce. There is a lack of open-source tools available for developers and researchers to take reliable energy measurements via hardware devices. In this paper, we present and demonstrate ARENA, a support tool that enables developers and researchers to connect to a physical measurement device without leaving the comfort of their IDE. Developers could use ARENA during development to compare energy consumption between different apps or versions of the same app. ARENA calculates energy consumption on an Android smartphone by executing a test scenario on the app under development. Further, ARENA helps aggregate, statistically analyze, report, and visualize the data, allowing developers and researchers to dig into the data directly or visually. We implemented ARENA as an IntelliJ and Android Studio plugin. A video demonstrating the usage of ARENA is available at \href{https://youtu.be/hgP5XL9SvRU}{\color{blue}https://youtu.be/hgP5XL9SvRU}.
\end{abstract}



\maketitle

\section{Introduction} \label{section-INTRODUCTION}

Portable devices such as mobile phones are popular. Over 1.9 billion mobile units were sold in 2018, and sales numbers are predicted to grow at a rate of 5\% every year \cite{Egham2018Gartner,Belkhir2018}. The constant increase in the use of mobile phones results in more energy consumption. Thus, one must consider the environmental impact and CO\textsubscript{2} footprint associated with it. Mobile phones are limited by their battery. Therefore, it is important that mobile apps are designed for energy-efficient usage. Based on recent studies, we know that user acceptance of energy-draining apps is low \cite{10.1145/3154384,chowdhury_greenscaler_2019,10.1145/3136014.3136031}. To make energy-efficient mobile apps, there is a need for tools that assist software practitioners in estimating the energy consumption of an app when it is running on a device. 

The energy measurement and analysis process typically involves setting up an energy measurement environment, executing the app under test (AUT) on the mobile device, and recording current/voltage data, usually at the rate of 5KHz and above. Once the energy data is acquired, it needs to be cleaned from noise and aggregated over several samples to account for variations in energy consumption due to background processes in the mobile device. Further, data is visualized or statistically analyzed to discover significant variations in energy consumption. The energy measurements could be captured either via hardware-based approaches (e.g., using devices such as Monsoon power monitor\footnote{https://www.monsoon.com/high-voltage-power-monitor}) or via software-based approaches (e.g., such as PowerAPI\footnote{https://github.com/powerapi-ng/powerapi-scala}). As compared to software-based approaches, hardware-based approaches are more accurate in capturing energy measurements but at the same time more cumbersome to implement. Several empirical studies exist \cite{Noureddine201221,KERN2018199,hindle_green_2015,Sahin201255} in which either one or both of these approaches are used to measure energy consumption of mobile apps. In each of these studies, the authors employ their own methods for measuring energy consumption, and most of the work is done manually or via specialized scripts. Therefore, it is difficult to compare and reproduce their results. Estimating the energy consumption of an Android app is challenging and resource-intensive. To overcome these problems, a systematic and fully/semi-automated process is needed to ensure that the measurements are performed consistently and reliably  \cite{MANCEBO2021106560}. Previously, many tools have been developed to estimate energy consumption \cite{6542365,10.1145/2451605.2451609,10.1145/2637364.2592003,7962354,10.1145/3038912.3052712,10.1145/2597073.2597097} of apps. However, they target large-scale app store analysis after an app has been published, or they use outdated hardware for physical measurements. Few tools exist that help developers estimate the energy consumption of an app during development. Android Profiler within the Android Studio IDE estimates energy consumption via a software-based approach but does not provide a means to analyze and report the energy consumption between different apps, or different versions of the same app, via a hardware-based approach. 
 
As the process of recording hardware-based energy measurement is lengthy with a steep learning curve. In this paper, we present an open-source tool \textbf{ARENA} (\textbf{A}nalyzing ene\textbf{R}gy \textbf{E}fficiency in a\textbf{N}droid \textbf{A}pps) to support the energy measurement and analysis process and to reduce the risks related to human errors. This tool integrates all the activities necessary to measure, statistically analyze and report (including result interpretations and visualization in the form of graphs) the energy consumption of Android apps.

In the rest of the paper, we describe ARENA's architecture and explain how ARENA supports the energy measurement and analysis process. Then, we provide implementation details. Finally, we present ARENA in a typical usage scenario and conclude the paper.

\begin{figure}
    \centering
    \includegraphics[width=\linewidth, height= 5cm]{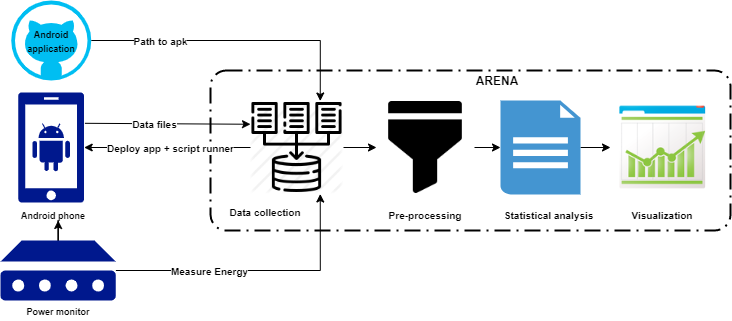}
    \caption{An overview of the energy measurement and analysis process that is supported by ARENA}
    \label{fig:1}
\end{figure}

\section{ARENA Architecture} \label{section-ARENA ARCHITECHTURE}

In this section, we describe how ARENA supports the energy measurement and analysis process. Typically energy measurement and analysis process consist of four steps: 1) Data collection, 2) Data pre-processing, 3) Statistical Analysis and 4) Visualization. For each of the main steps in this process, there exists a corresponding component in ARENA. Fig. 1 gives an overview of the energy measurement and analysis process that is supported by ARENA.   

 \subsection{Component 1: ExperimentRunner}

The ARENA component \lq ExperimentRunner' supports the first step of the energy measurement and analysis process, i.e., \lq Data collection'. During data collection, energy measurements are recorded several times to account for the underlying variations in energy consumption due to background processes in the mobile device. The energy measurements are controlled from the host computer via \lq ExperimentRunner', which helps ARENA user to control activities required to set up and execute the experiment for data collection. \lq ExperimentRunner' initializes the callback object to display output in the tool window. It also creates a directory where customized shell scripts will be created.

\subsubsection{Experiment Setup}

Before using the ARENA tool, the scope of the experiment, measurement environment settings, the AUT, and test scripts are prepared by the ARENA user. R version 3.4.3 or above and Rtools needs to be installed on the host computer. To measure hardware-based energy readings, ARENA is designed to be used with the Monsoon Power Monitor (MPM), a popular physical measurement device that has been used in various studies \cite{8870158,10.1145/2597073.2597085,7884613}. Therefore related libraries and Python packages\footnote{https://figshare.com/s/9cdfc9f8b39411698afd} need to be installed as per the user manual of MPM. The mobile phone on which the AUT will be executed should have Android 5.0 and above. The mobile phone is connected to the host computer with a USB cable via MPM disabling the USB phone charging once the energy measurement starts. ARENA user should check that the screen brightness is set to a minimum and only essential Android services are running on the phone. \lq ExperimentRunner' checks if the mobile device is successfully connected to the host computer. Additionally, a script \lq start\_power\_monitor.py' is generated, which initializes MPM runtime current limits, and enables the USB channel on MPM hardware. The serial number of MPM hardware can be configured in the \lq start\_power\_monitor.py' script in the ARENA source code. \lq ExperimentRunner' works with only one MPM at a time.

\subsubsection{Experiment Execution}

\lq ExperimentRunner' creates customized scripts for experiment execution. These scripts include commands to clear battery statistics, memory statistics, network statistics and adb log files before each iteration of the experiment. We consider an iteration to be the execution of a test case on a mobile device once. Based on the AUT selected by the ARENA user, commands in shell scripts are customized to install and run the app (for baseline readings, these commands are not included). The script \lq start\_power\_monitor.py' creates an instance of the sample engine class from the MPM Python library. By default, the current/voltage samples are saved as a Python list that can be retrieved with the getSamples() function. At the end of each iteration, the Python list is converted into a CSV file and saved on the host computer. MPM output files are checked for reliability based on the number of dropped samples. As MPM records samples at a rate of 5KHz, and assuming that AUT runs for more than one second, the ARENA user is given a warning to check current/voltage data if 1000 or more samples are dropped.

Once all scripts are ready, they are pushed to the mobile device along with the script-runner apk. Script-runner is a small app that comes with ARENA to automatically trigger AUT and related shell scripts on the mobile device. This app is a necessary overhead to save manual effort and to ensure that no problems are created during the experiment due to human error. \lq ExperimentRunner' gives the options to the ARENA user to re-run the same iteration, run the next iteration, uninstall AUT from the mobile device, and clear data for AUT from the mobile device. The selected option is passed as a runtime argument to the script-runner app, which executes the relevant shell script on the mobile device.

After each iteration, data is retrieved from the device to the result folder on the host computer, and files are renamed as per iteration number. e.g. for the first iteration, the adb log file "logcat.txt" is renamed to "Logcat\_R1" and so on. During the next iterations, settings are updated in the shell scripts (if needed).
 
 \subsection{Component 2: CleanupRunner }
 
The ARENA component \lq CleanupRunner' supports the second step of the energy measurement and analysis process, i.e., \lq Data pre-processing'. \lq Cleanup-Runner' renders the raw data files in a list in the tool window and performs cleaning/filtering on the selected files. PID (process-ID) and UID (user-ID) are extracted from the adb log files for each iteration of the experiment. The UID is used to extract relevant data from network statistics files. CPU and memory statistic files are filtered by app package name. For cleaning adb logs, UID, PID, and user-specified tags are used. As the format of adb log and statistic files in different Android versions is slightly different, to produce cleaned output files with a consistent format, the API version of the mobile device is used to implement the correct parser on the log and statistic files. Once the adb log and statistic files are cleaned, the timestamps from the cleaned adb log file are used to extract relevant current/voltage data in each iteration. The cleaned current/voltage file is used to calculate energy consumption in joules (J) of AUT in each iteration. An average of baseline energy is subtracted from the calculated energy consumption of AUT (under the assumption that an increase in energy consumption from the baseline is due to the execution of AUT). A data file named \lq data.csv' is created containing the package name of AUT, energy (J), memory \%, CPU \%, and network statistics for each iteration. Another file named \lq average\_data.csv' is created with aggregated values for energy (J), memory \%, CPU \%, and network statistics of all iterations of AUT.
 
 \subsection{Component 3: AnalysisRunner}

The ARENA component \lq AnalysisRunner' supports the third step of the energy measurement and analysis process, i.e., \lq Statistical Analysis'. \lq AnalysisRunner' populates the combo boxes for dependent, independent and filter variables in the tool window with column names from the selected CSV data files (the cleaned energy files produced by \lq CleanupRunner' are used here). \lq AnalysisRunner' provides detailed help text in the tool window to make it easier for the user to select a statistical analysis based on requirements and data type. Based on the ARENA user selection, the values of variables included in the analysis are updated in the relevant R scripts, which are then executed to produce a report containing the results of the selected statistical analysis and its interpretation.
 
 \subsection{Component 4: VisualizationRunner}
 
The ARENA component \lq VisualizationRunner' supports the fourth step of the energy measurement and analysis process, i.e., \lq Visualization'. \lq VisualizationRunner' populates the combo boxes for dependent, independent, and filter variables in the tool window with column names from the selected CSV file. The type of the graph selected and the dependent, independent, and filter variables control how the data in the graph is displayed. \lq VisualizationRunner' allows various graph configurations for each graph type in terms of label font, legend colours, graph title, graph size, sequence of labels on the x-axis, etc.

\section{ARENA Implementation}

ARENA is built for integration with IntelliJ IDEA and Android Studio IDE as a plugin. Functionalities of the plugin are implemented on widgets of the tool window (from here onwards referred to as ARENA interface). The plugin is implemented in Java. Each component in ARENA's architecture corresponds to a tab on the ARENA interface. Based on the scope and requirements of the experiment, the ARENA user can set certain parameters on each tab to get the results. It is ideal to use the tabs in the ARENA interface iteratively as they are interrelated. However, if ARENA users want to reuse data of a particular process step or skip a process step, that is also permitted. 

The first tab in the ARENA interface is \lq Data collection', which corresponds to the ARENA component \lq ExperimentRunner'. Within this tab, ARENA users can perform two sets of activities, 1) configure experiment setup by selecting energy measurement mode, data collection phase and corresponding data files, and 2) control the experiment execution by configuring the experiment parameters such as number of iterations (a single iteration is the execution of test apk once, the choice of this value depends on the requirement of the experiment, however for the sake of sampling distribution a value between 10-30 is considered good), path to app apk, path to test apk, data path on mobile device, test class, test runner, re-run configuration (i.e., re-install app or clear data), results folder etc. The main output of this tab is the raw current/voltage data from MPM and corresponding adb logs and statistics from the mobile device. 

The second tab in the ARENA interface is \lq Pre-processing', which corresponds to the ARENA component \lq CleanupRunner'. The main outputs of this tab are the 1) filtered current/voltage data, adb logs and statistics\footnote{Details of columns in filtered data file https://figshare.com/s/50c5732300315023b197}, and 2) calculated and aggregated energy and statistics data\footnote{Details of columns in aggregated data file  https://figshare.com/s/cbc2fd529b413e4dcbf1}.

The third tab in the ARENA interface is \lq Analysis', which corresponds to the ARENA component \lq AnalysisRunner'. The main output of this tab is a report(s) in .docx format with the results of statistical analysis about the energy consumption of AUT (along with its interpretations).

The fourth tab in the ARENA interface is \lq Visualization', which corresponds to the ARENA component \lq VisualizationRunner'. The main output of this tab is the graph of the selected type.

In all the tabs, hovering the mouse pointer on a widget of the interface shows a tool-tip with help text. Progress and error messages are shown either in the tool window terminal or via error labels.
  
\section{ARENA in practice} \label{section-ARENA USAGE}

This section ties together all components described above and provide example usage. ARENA can be installed as an IntelliJ or Android Studio plugin using the package we provide on our bitbucket repository\footnote{https://bitbucket.org/hinaanwar2003/arena/src/master/ The plugin is packaged as a zip file EnergyPlugin-1.0-SNAPSHOT.zip }. After installation, when a user opens a new or existing project, they can see the ARENA tab on the right side of the IDE. 

\subsection{Typical Usage Scenario}

A typical usage scenario of ARENA begins with the source code of the AUT. The developer writes automated Android user interface tests for AUT using tools such as Espresso\footnote{https://developer.android.com/training/testing/espresso}. Next, the developer wants to assess AUT's energy consumption to compare with the previous version of the same app or against a competitor app. The ARENA interface facilitates the developer to measure, aggregate, analyse, and visualize the energy consumption of AUT. Using the \lq Data collection' tab, the energy data collection process is initiated. The corresponding adb logs and additional statistics (if selected), such as CPU, memory, network statistics, and trace files, are recorded and extracted from the mobile device. The energy data from MPM is automatically saved as a CSV file on the host computer. Using the \lq pre-processing' tab, the raw energy data is cleaned and aggregated by matching it against the start and end timestamps found in the adb log files. Using the \lq Analysis' tab, various statistical analyses (such as Summary statistics, Kruskal-Wallis, Spearman Correlation, ANOVA, etc.) could be performed on the data. After the analysis is complete, a detailed report of the analysis and the interpretation of the results is generated (in .docx format). Using the \lq Visualization' tab, the data could be visualized by creating various graphs (such as scatter plot, box plot). In Fig. 2, we show the detailed workflow with included sub-steps supported by the ARENA tool. For the upper part of the workflow (labelled \lq Data collection'), Fig. 3 shows the corresponding ARENA interface\footnote{See the detailed tool tutorial: https://figshare.com/s/4c4ec26fc0ec91fbad41} in IntelliJ IDE.

\begin{figure}[h]
	\centering
	\includegraphics[width=\linewidth, height=8cm]{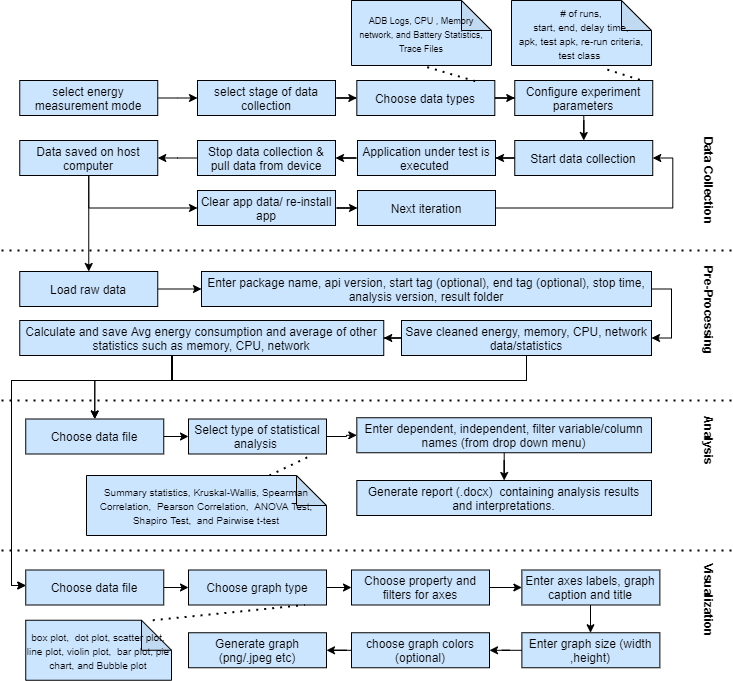}
	\caption{Detailed workflow supported by ARENA tool}
	\label{fig: Workflow explaining ARENA's interface}
\end{figure}

\begin{figure}[h]
	\centering
	\includegraphics[width=\linewidth, height= 7cm]{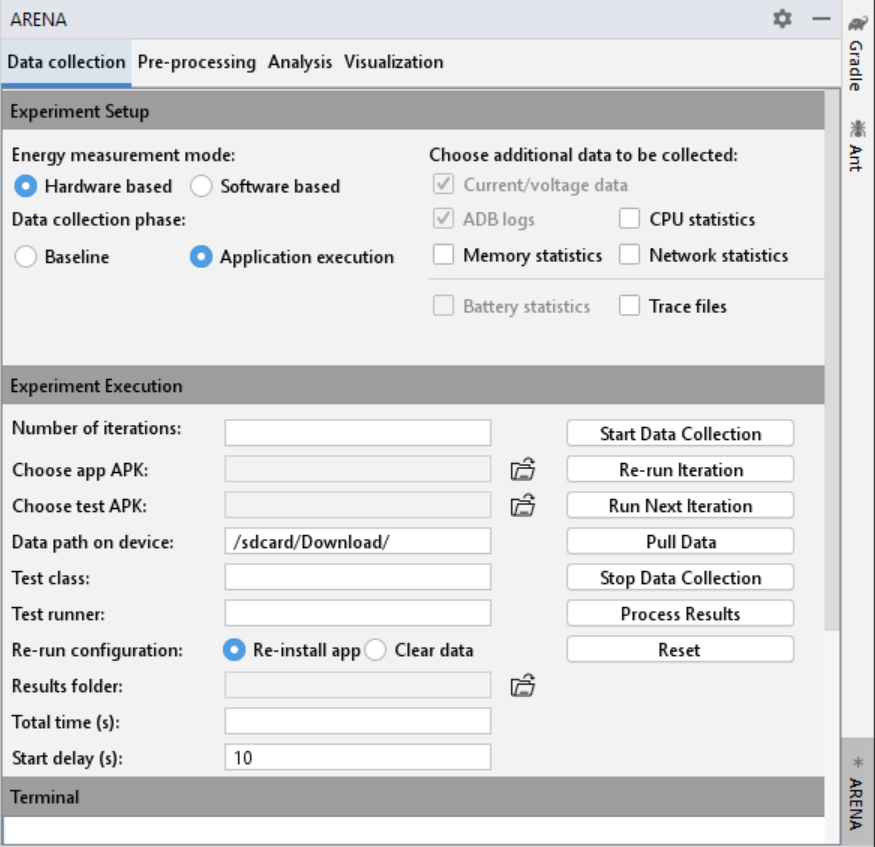}
	\caption{ARENA interface}
	\label{fig:ARENA's interface}
\end{figure}

\subsection{ Application Example}

We used ARENA in a study \cite{10.1145/3387905.3392095} to evaluate the energy consumption of commonly used third-party network libraries in Android apps. We made 45 versions of a custom app using selected third-party network libraries in different use cases. We used ARENA to measure the energy consumption of each version of the custom app by executing it on an Android device ten times. We recorded 450 energy measurements along with corresponding adb logs. Next, the recorded data was cleaned, aggregated, analyzed, and visualized using ARENA to identify the statistically significant changes in energy consumption of different third-party network libraries in different use cases. ARENA significantly reduced the time and effort required to measure and analyze the energy consumption of AUT. As the process described was controlled via ARENA, errors in measurement due to human error were also avoided. 

\section{Conclusion} \label{section-CONCLUSION}

This paper presents ARENA, a support tool for developers and researchers to compare energy consumption between versions of the same app or different apps. The energy consumption of the app can be measured using software or hardware-based approaches. Compared to software-based approaches, hardware-based approaches for collecting energy data are more accurate but difficult to apply. ARENA connects with one of the most widely used physical measurement devices (Monsoon Power Monitor) to capture energy data. ARENA provides an interface that is consistent with the IntelliJ and Android Studio IDEs. Furthermore, ARENA facilitates the aggregation, statistical analysis, reporting, and visualization of data. The implementation of ARENA is available at \href{https://bitbucket.org/hinaanwar2003/arena}{\color{blue}https://bitbucket.org/hinaanwar2003/arena}.

\begin{acks}
This work is supported by the Estonian Center of Excellence in ICT
research (EXCITE) and the Estonian state stipend for doctoral
studies. 
\end{acks}

\bibliographystyle{plain}
\bibliography{reference}

\end{document}